\documentclass[,linenumbers]{aa}
\usepackage{graphicx}
\usepackage[breaklinks,colorlinks,urlcolor=blue,citecolor=blue,linkcolor=blue]{hyperref}
\usepackage{multirow}

\usepackage{txfonts}

\newcommand\aastex{AAS\TeX}
\newcommand\latex{La\TeX}
\newcommand{\lr}{$\mathrm{L_R}$}

\newcommand{\lx}{$\mathrm{L_X}$}
\newcommand{\vcme}{$\mathrm{V_{CME}}$}
\newcommand{\rec}{$\mathrm{\phi_{rec}}$}

\newcommand{\Pfl}{$\sqrt{\mathrm{L_X\phi_{rec}}}$}
\newcommand{\Pcme}{$\sqrt{\mathrm{L_RV^2_{CME}}}$}
\newcommand{\Pf}{$\mathrm{P_{flare}}$}
\newcommand{\Pc}{$\mathrm{P_{CME}}$}
\newcommand{\Cor}[2]{C$_\mathrm{#1,#2}$}
\newcommand{\crc}{\Cor{R}{C}}

\begin{document}

\title{Novel scaling laws to derive spatially resolved flare and CME parameters from sun-as-a-star observables}

    \titlerunning{Novel scaling laws for solar-stellar ``flare-CME-type II'' events}

   \author{Atul Mohan\inst{1,2}
         , Natchimuthuk Gopalswamy\inst{1}
         , Hemapriya Raju\inst{3,1}
         \and
         Sachiko Akiyama\inst{1,2}
         }
        \institute{
    NASA Goddard Space Flight Center,  8800 Greenbelt Road, Greenbelt, MD, 20771, USA
    \and
    The Catholic University of America, 620 Michigan Avenue, N.E. Washington, DC 20064, USA
    \and
    Indian Institute of Technology, Indore, Simrol, Indore, 453552, India
    \and
    \email{atul.mohan@nasa.gov}
    }
    \authorrunning{Mohan, A. et al.}

   \date{Received June 10, 2024; Accepted September 27, 2024}
\abstract
{Coronal mass ejections (CMEs) are often associated with X-ray (SXR) 
flares powered by magnetic reconnection in the low corona, while the CME shocks in the upper corona and interplanetary (IP) space accelerate electrons often producing the type II radio bursts.
The CME and the reconnection event are part of the same energy release process as highlighted by the correlation between reconnection flux (\rec) that quantifies the strength of the released magnetic free energy during the SXR flare, and the CME kinetic energy that drives the IP shocks leading to type II bursts. 
Unlike the Sun, these physical parameters cannot be directly inferred in stellar observations.
Hence, scaling laws between unresolved sun-as-a-star observables, namely SXR luminosity (\lx) and type II luminosity (\lr), and the physical properties of the associated dynamical events are crucial.
Such scaling laws also provide insights into the interconnections between the particle acceleration processes across low-corona to IP space during solar-stellar "flare-CME-type II" events.
Using long-term solar data in the SXR to radio waveband, we derived
a scaling law between two novel power metrics for the flare and CME-associated processes. The metrics of "flare power" (\Pf=\Pfl) and "CME power" (\Pc=\Pcme), where \vcme\ is the CME speed, scale as \Pf $\propto$ \Pc$^{0.76\pm0.04}$. In addition, \lx\ and \rec\ show power-law trends with \Pc\ with indices of 1.12$\pm$0.05 and 0.61$\pm$0.05, respectively. These power laws help infer the spatially resolved physical parameters, \vcme\ and \rec, from disk-averaged observables, \lx\ and \lr\, during solar-stellar flare-CME-type II events.}

\keywords{stars: flare - stars: activity - stars: magnetic field - stars: coronae - radio continuum: stars - stars: low-mass
               }
 \maketitle              
\section{Introduction} \label{sec:intro}
Coronal mass ejections (CMEs) are a major driver of space weather events in solar and stellar environments~\citep[see,][for an overview]{howard23_CME_review}.
A CME can produce shocks across corona out to interplanetary (IP) space, accelerating electrons which often produce meter-kilometer~(m-km) band type II bursts~\citep{wild50_bursttypes,wild1970,Mclean85_book,Gopal11_PREconf,Miteva17_SEP-radburst_link,alvarado22_CME_star}.
Depending on the heights of shock formation, the type II bursts may have a start and end frequency anywhere within the m-km range.
The CMEs associated with the type II bursts are generally much faster and wider than an average CME, with 78\% of them associated with major solar energetic particle (SEP) events~\citep{gopal05_typeIIs_SEPs,anshu23_typeII_stats}.
The type II associated CMEs are also accompanied by relatively intense soft X-ray (SXR) flares~\citep{gopal06_CMEtypeII_book}. Hence, the type II bursts are highly sought after in active stars as a CME shock signature~\citep[e.g.,][]{Osten08_ADLeoFinebursts,2018ApJ...856...39C,Villadsen14_First_detect_SLS_inRadio_VLA}. 

When the type II burst conveys the strength of particle acceleration caused by the CME shocks, the associated SXR flare is produced by the heating of the post-reconnnection magnetic field structures in the low corona.
Hence the peak SXR luminosity (\lx) correlates well with the reconnection flux (\rec), which is a proxy for the strength of the reconnection-driven currents that drive the local heating~\citep{Kazachenko17_Ribbondb,Sindhuja20_Recflx_xflr}.
{\cite{gopal18_coronalRcflx_IPburstconnection} show that \rec\ correlates well with CME kinetic energy, revealing a link between the low-coronal and IP space impacts of the energy release event, of which the CME and the flare are a part (hereafter, flare-CME-type II event).
However, there exist no scaling laws involving \lx, type II luminosity (\lr), and the physical properties of the flare and the CME, which could reveal any links between the mechanisms and drivers of particle acceleration across corona to IP space.
In addition, since the flare and type II burst outshine the quiet sun during a flare-CME-type II event, \lx\ and \lr\ are sun-as-a-star observables. So, the scaling laws of the aforementioned kind are relevant for stellar CME studies, which lack spatially resolved observations.
In this context, the G\"uedel-Benz relationship~\citep[GBR;][]{guedel93_dM_GBZ} that connects \lx\ and microwave (5 - 8\,GHz) luminosity by a power law is noteworthy. GBR revealed the link between the population of flare-accelerated electrons propagating toward the Sun causing the microwave and SXR flare, and it provides a common framework to explore the solar and stellar flares~\citep{guedel93_GBR_F-M,guedel95_GBR_FGgiants,airapetian98_GBRtheory}.
This work explores analogous correlations between \lx, \lr, and the physical properties of the flare-CME-type II events.}

Section~\ref{sec:data} presents the data sources and the event catalog. The analysis results and their interpretation are described in Sect.~\ref{sec:results}, followed by conclusions in Sect.~\ref{sec:conclusion}.
\section{Data and methodology} \label{sec:data}
We used the calibrated decameter-hectometric (DH) radio dynamic spectra from Radio and Plasma Wave Investigation (WAVES) instruments on board Wind, STEREO A, and STEREO B spacecraft. 
The DH type IIs were chosen for this study over the metric bursts for multiple reasons.
Firstly, the DH type II bursts are produced at regions of CME-driven interplanetary shocks caused by relatively stronger flares and faster CMEs than their metric counterparts~\citep{Gopal11_PREconf,Miteva17_SEP-radburst_link}. 
Also, DH bursts are more closely associated with SEPs and sustained $\gamma$-ray emission than the type IIs confined within the metric band, making these bursts more interesting from a space weather perspective~\citep{gopal06_CMEtypeII_book,Miteva17_SEP-radburst_link,gopal18_SGRE-typeIIlink}.
The database of the calibrated DH dynamic spectra from each spacecraft gathered over multiple solar cycles forms a uniformly calibrated long-term dataset.
In addition, simultaneous STEREO and Wind observations form a unique multi-vantage point radio database. 
 
\begin{figure}
\centering
\includegraphics[width=0.45\textwidth,height=0.4\textheight]{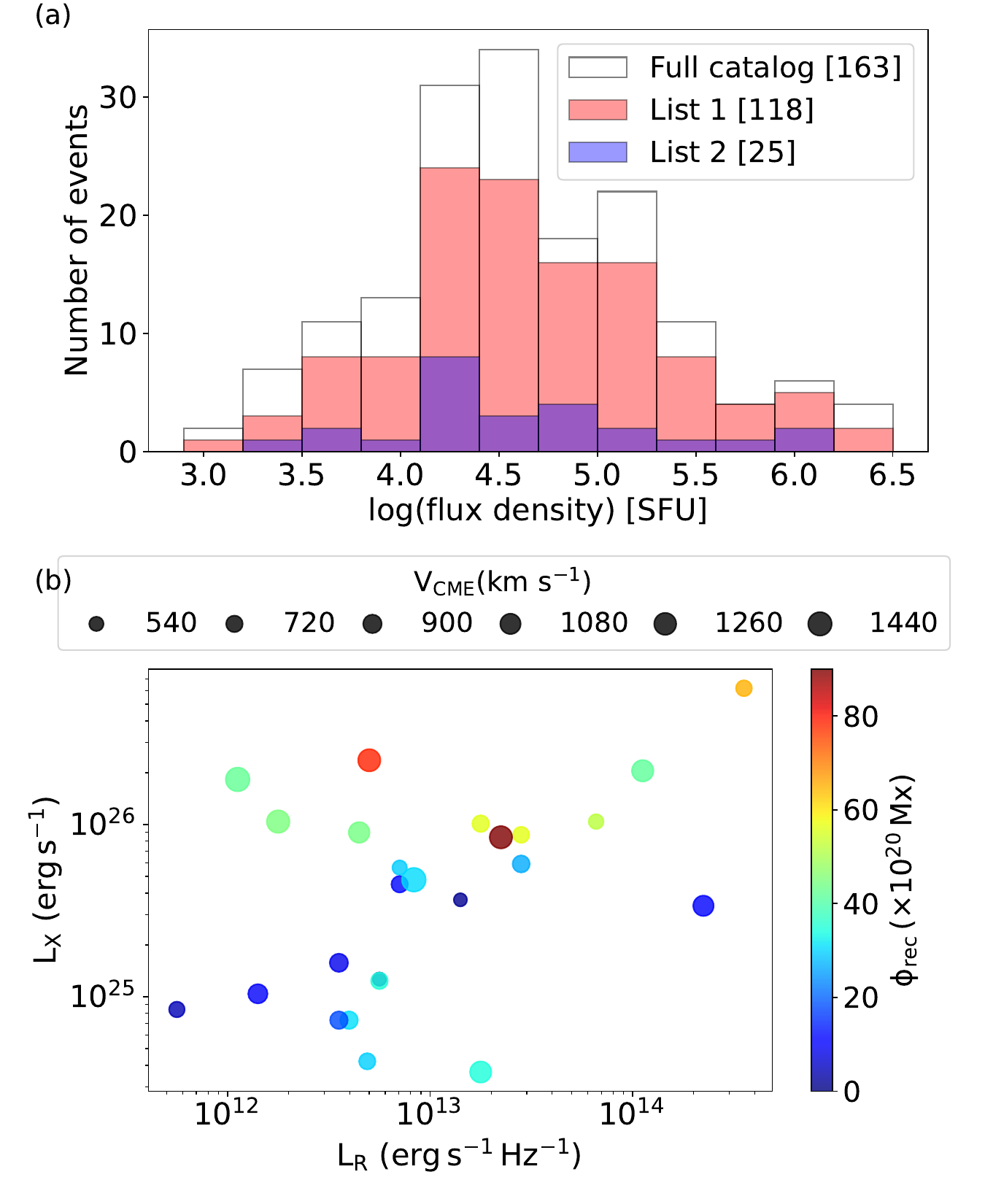}
\vspace*{-2mm}
\caption{Statistical properties of flare-CME-type II events. (a) The \lr\ histogram for DH type II event lists and the full catalog. Sample sizes are given in square brackets. The distributions look similar across event lists with a median flux density of $10^{4.5}$\,SFU. (b) Properties of List 2 events. The \lx\ is not well correlated with \lr.}
\label{fig1:result}
\vspace{-0.2cm}
\end{figure}

The events reported in the DH type II catalog~\citep{Gopal19_DHtypeII_catalog} between November 2006 and July 2023 constitute the initial sample. For each event, data from all spacecraft were examined and the best-recorded event dynamic spectrum (DS) was chosen for further analysis. The peak flux of the burst within the 3 - 7\,MHz range was used to compute \lr. Within 3 - 7\,MHz the major burst types include type II and type III bursts. The long-duration broadband type-IV bursts are mostly confined above 7\,MHz~\citep{Atul24_DHtypeIVcatalog}. {Type IIIs have very high frequency-drift rates in comparison to type IIs,} making it easier to disentangle the emission from near-simultaneous bursts in the DS. As a result, the 3 - 7\,MHz range is an ideal choice to ensure robust flux estimation in most cases with minimal impact from co-temporal bursts. The 
\lx\ and CME parameters were obtained by cross-matching the radio event list with the publicly available catalogs maintained by the Coordinated Data Analysis Workshops (CDAW) group\footnote{\href{https://cdaw.gsfc.nasa.gov/}{https://cdaw.gsfc.nasa.gov/}}. The resultant catalog (hereafter "full catalog") has 163 bursts with 3- 7 \,MHz band \lr\ estimates. Of these, 118 events (List 1) have a good estimation for \lx\ and CME properties, especially their speed (\vcme). 
From List 1, the type II bursts that are well isolated in the DS, without potential contamination from other emission features, were chosen. The \rec\ was estimated for all possible events using the post-eruption arcade (PEA) method described in \cite{Gopal17_FREDtech}. 
All of these resulted in a list of 25 flare-CME-type II events (List 2) with reliable estimates for \lx, \lr, \vcme, and \rec, which are the focus of our analysis.
{ The full catalog, event lists, the calibrated DS across the entire WAVES band and within 3 - 7\,MHz, and a detailed description of the tables are made available online\footnote{\href{https://cdaw.gsfc.nasa.gov/CME_list/radio/multimission\_type2/}{https://cdaw.gsfc.nasa.gov/CME\_list/radio/multimission\_type2/}} and via Zenodo\footnote{\href{https://doi.org/10.5281/zenodo.13896427}{https://doi.org/10.5281/zenodo.13896427}}.}

\section{Results} \label{sec:results}
\begin{figure*}
\centering
\includegraphics[width=0.9\textwidth,height=0.5\textheight]{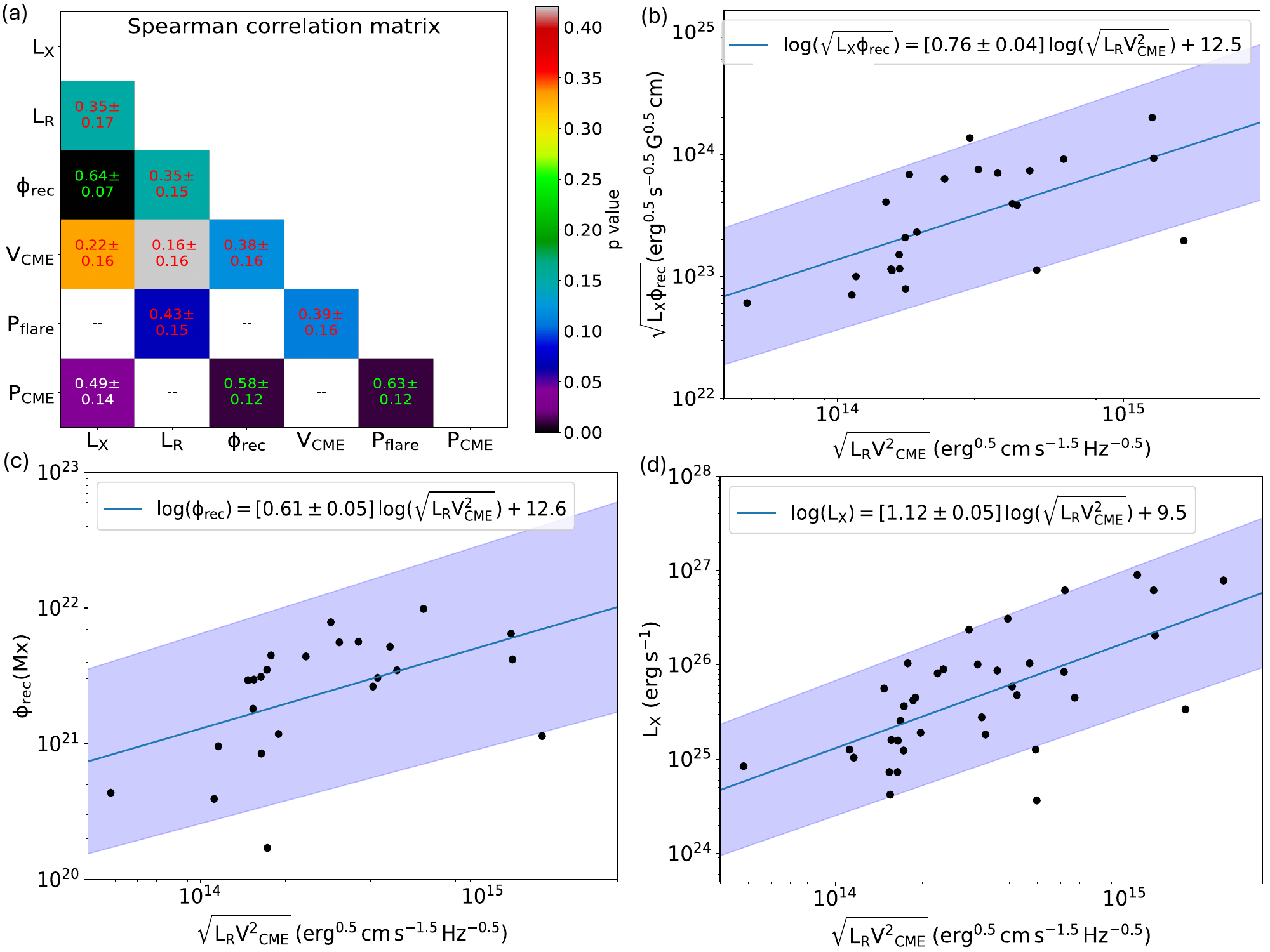}
\vspace*{-2mm}
\caption{Cross-correlations and scaling laws between various properties of flare-CME-type II events. (a) Spearman correlation matrix. 
Each cell provides the Spearmann cross-correlation coefficient (\crc) between the respective row (R) and column (C) parameters. The cell color represents the p value. The \crc, error, and p-value were found using a Bootstrapping analysis. The "very significant"~(p$<$0.015), "significant"~(p$<$0.05), and "insignificant"~(p$>$0.05) \crc\ values are shown in green, white, and red text, respectively. The \crc\ between the novel composite metrics and the respective dependent parameters are masked and marked by "--". (b) The \Pf - \Pc\ scaling law. (c-d) Relations connecting \rec\ and \vcme\ to \Pc. Shaded regions represent the fitting uncertainty intervals.}
\label{fig2:result}
\vspace{-0.2cm}
\end{figure*}
\subsection{Statistics and cross-correlation analysis}
Figure~\ref{fig1:result}a shows the histogram of \lr\ for the full catalog and the sub-lists. The median flux density is $\sim 3.5\times 10^4$\,SFU for all lists. Figure~\ref{fig1:result}b shows the flare and CME properties of List 2 events. There is no significant correlation between \lr\ and \lx. However, certain other parameter pairs appear to be correlated, which are subsequently discussed in detail.
Two novel metrics of flare power (\Pf = \Pfl) and CME power (\Pc = \Pcme) are defined by combining \rec\ and \vcme, with \lx\ and \lr\,, respectively. 
{ Spearman cross-correlation analysis was carried out between each of the parameters considered, namely \lx, \lr, \rec, \vcme, \Pf\ and \Pc. 
Given the small sample size, a bootstrap technique was employed to obtain robust correlation coefficients and p values.
Of the 25 events, 18 ($\sim$70\%) were randomly selected iteratively while computing cross correlations, \crc, between each parameter pair (R,C) over 10000 rounds. The median of the 10000 \crc\ and p values for each pair were noted. The errors on each median \crc\ is the median absolute deviation. Fig.~\ref{fig2:result}a shows the cross-correlation matrix with each cell providing the median \crc\ between the respective row (R) and column (C) parameters.
}
The cell color denotes the median p value, and the text color of \crc\ denotes the statistical significance. We note that{ 
$|$\crc $|$ values with p$<$0.015 are considered very significant (green text), while 0.015 $<$p$<$0.05 are significant (white text), and those with p$>$0.05 are insignificant (red text).  
}
The \crc\ for the tuples (\Pf,\lx), (\Pf,\rec), (\Pc,\lr), and (\Pc,\vcme) have been ignored (marked by "--") since the row parameter is a function of the column parameter by definition. 

{ The parameter tuples with a significant correlation include (\rec,\lx), (\Pc,\Pf), (\Pc,\lx), and (\Pc,\rec).}
Our $\mathrm{C_{L_X,\phi_{rec}}}$ matches the recent result by \cite{kazachenko23_Ribbondb}.
However, both \lx\ and \rec\ are related to the low-coronal flare, while we aim to identify links between the powering of the low-coronal flare and the interplanetary type II burst. This makes the other significantly correlated tuples more interesting.

\subsection{Flare-CME-type II scaling laws} \label{sec:modGBR}
Of all possible parameter pairs where one is purely related to the SXR flare and the other to radio burst physics,{ the engineered novel power metrics} shows the highest correlation.  
We derived a new robust scaling law between the powers in the flare- and CME-driven activity,
\begin{eqnarray}
    \sqrt{\mathrm{L_X\phi_{rec}}} = (1 \pm 8.5)\times 10^{12.5}\left(\sqrt{\mathrm{L_RV^2_{CME}}}\right)^{0.76\pm0.04}.
    \label{eqn:GBRm}
\end{eqnarray}
{Interestingly, the power-law index coincides with that of GBR~\citep[\lx\ $\propto$ \lr$^{0.73}$;][]{guedel95_GBR_FGgiants} within estimation errors. GBR computes \lr\ in the 5 - 8\,GHz range.}
The other significant, strongly correlated tuples also provide useful scaling laws;
\begin{eqnarray}
    \mathrm{L_{X}} &=& \mathrm{ (1 \pm 10)\times 10^{9.5}\left(\sqrt{\mathrm{L_RV^2_{CME}}}\right)^{1.12\pm0.05}}\\
    \label{eqn:vcme}
    \mathrm{\phi_{rec}} &=& (1 \pm 10.4)\times 10^{12.6} \left(\sqrt{\mathrm{L_RV^2_{CME}}}\right)^{0.61\pm0.05}.
    \label{eqn:rec}    
\end{eqnarray}
We note that \vcme\ and \lr\,, which make up \Pc, have insignificant or no correlation with \lx\ and \rec, making the strong \Pc - \Pf\ correlation robust.
{The power-law indices are constrained fairly well. More data from the ongoing solar cycle will help better constrain both the fit parameters in the future.}
\vspace{-0.3cm}
\section{Discussion}
{ 
\subsection{\Pc\ - \Pf\ scaling law: Physical insight}
The flare-CME-type II phenomenon has two major sub-events, the "reconnection-heating" event in the low corona and the "CME - particle acceleration" event extending the impact of the energetic phenomenon to higher corona and IP space leading to various space weather effects.
The populations of accelerated electrons and the powering mechanism are different in both sub-events, though the evolution of the sub-events is causally and physically connected.
As a result, robust scaling relations that connect the various observable features of the sub-events of a flare-CME-type II phenomenon remained elusive (see, Sec.~\ref{sec:intro}).
However, our physics-informed "feature engineering"~\citep{ML_book} provides novel metrics, \Pc\ and \Pf, with a significant correlation between them and with other parameters (Eqns.~\ref{eqn:GBRm}--~\ref{eqn:rec}). The
\Pc\ and \Pf\ are defined as proxies for the mean power in each sub-event using direct observables: \Pf\ for the reconnection-heating event and \Pc\ for the CME-particle acceleration event.
The reconnected magnetic flux (\rec) leads to strong local electric fields accelerating electrons in the coronal post-flare loops leading to local heating and associated thermal SXR flares. So \lx\ is a direct proxy for the heating power. Meanwhile, \rec\ is a direct observable proxy for the power in the reconnection-driven electric fields. We note that \Pf\ estimates the coupled power in the reconnection and heating, providing a proxy for the power in the reconnection-heating event. 
Meanwhile, the erupted CME following the reconnection event generates a propagating shock, accelerating electrons and causing the type II burst.
The direct observable proxy for the CME shock power is the \vcme$^2$ and \lr\ is the proxy for the particle acceleration strength.
Similar to \Pf, \Pc\ estimates the typical power of the CME-particle acceleration event.
The novel power metrics, \Pf\ and \Pc, thus couple the strength of local particle acceleration with the power in their respective drivers for the two major sub-events of a flare-CME-type II phenomenon.
}
\vspace{-0.5cm}
\subsection{Comparison with GBR}
{ The \Pc - \Pf\ scaling law can be philosophically compared to the GBR. GBR relates the power in the low coronal particle acceleration (\lr$_{\rm ,5 - 8\,GHz}$) and subsequent heating (\lx). Any attempted extension or a search for a similar \lr-\lx\ scaling has not quite been successful in the meterwave band~\citep[e.g.,][]{callingham21_noGBR_LoTSS,Yiu24_noGBR_vlass}.
A major physical reason for this is that the metric bursts have a variety of dynamic spectral types, with each type having a unique driving mechanism. For this work we chose the type II bursts since they are highly sought after in solar and stellar CME studies. Based on the power in the different drivers of the radio and SXR emission in flare-CME-type II phenomena, we modified the \lr\ term in GBR to \Pcme\ and the \lx\ to \Pfl.
Interestingly, the new power terms show a power-law index similar to the GBR, while also accounting for the differing SXR and radio emission formation scenarios~\citep{guedel95_GBR_FGgiants}. A similar careful analysis for other radio burst types may also provide meaningful correlations involving \lr\ and \lx.
}
\subsection{Implications to stellar CME studies}
{
Being disk-averaged, \lx\ and \lr\ are observable in other stars. 
Combining these observables and the scaling laws associated with the strong correlations (Eqn.~\ref{eqn:GBRm}, and Eqn.~\ref{eqn:rec} or the \lx-\rec\ scaling law estimated over a larger sample~\citep{petsov03_Lx_recflx_solar-stellar}), the unknown \vcme\ and \rec\ can be estimated. We note that \vcme\ and \rec\ cannot be directly inferred in stars since flare imaging observations are impossible.
In addition, since the type II emission forms primarily via a plasma emission mechanism, the observation frequency ($\nu$) relates to the local electron density (n$_\mathrm{e}$; $\nu \propto \sqrt{\mathrm{n_e}}$)~\citep{ginzburg1958,melrose1970}, the type II frequency drift rate ($\mathrm{\delta \nu/\delta t}$) and the \vcme\ estimate can provide the density gradient ($\mathrm{\nabla_h n_e}$) across the coronal height (h):
\begin{eqnarray}
    \mathrm{\frac{\delta \nu}{\delta t}} &=& \mathrm{\frac{\delta \nu}{\delta n_e}\frac{\delta n_e}{\delta h}\frac{\delta h}{\delta t}} \\
    \mathrm{\frac{\delta \nu}{\delta t}} &=& \mathrm{\frac{\nu}{2n_e}\nabla_h n_e V_{CME}},  
\end{eqnarray}
where \vcme\,=$\mathrm{\delta h/\delta t}$. Furthermore, \vcme,\, \rec\,, and $\mathrm{\nabla_h n_e}$ are crucial constraints to flare and CME evolution models. 
}

{
Since DH type IIs and metric bursts have a similar emission mechanism, the results presented provide a direction to extend the \lr - \lx\ correlation studies to the metric band by carefully considering the physical drivers of SXR and radio emission. The extension of the analysis presented here to the metric waveband is beyond the scope of this work given the lack of a uniformly calibrated long-term metric dynamic spectral database, unlike the DH band.
However, the results presented provide a scientific motivation to calibrate and explore the metric type II events using a similar methodology.
However, one may have to use the CME speed in the lower coronal heights probed by the meterwave band as opposed to the IP space \vcme\ used here. 

Given the low occurrence rate of CME-associated metric bursts in active stars~\citep{2019ApJ...871..214V,Zic20_typeIV_ProximaCen,atul24_ADLeotypeIV}, and the expected higher occurrence rate in the DH band~\citep{alvarado22_CME_star}, there has been a push for space-based DH band observatories.
Our results aid this initiative by providing the flux estimates of the flare-CME-type II events with varying \vcme - \rec\ values.  
}
\section{Conclusion} \label{sec:conclusion}
We present a novel scaling law for flare-CME-type II events using decades-long solar decameter-hectometric type II burst data and the simultaneous multi-waveband imaging and non-imaging database.
The scaling relation links two novel metrics of flare power (\Pf = \Pfl) and CME power (\Pc = \Pcme), where \lx\ is the SXR luminosity, \lr\ is the type II luminosity, \rec\ is the reconnection flux, and \vcme\ is the CME speed. When \Pf\ encapsulates the power in low-coronal heating and the associated SXR flare, \Pc\ relates to CME shock-driven particle acceleration and the associated type II burst. The two terms together encapsulate the two major manifestations of the flare-CME-type II event in the low-coronal (reconnection-heating) and interplanetary (CME-particle acceleration) regions. We note that
\Pf $\propto$ \Pc$^{0.76}$.
Analogous to the GBR, which links the flare thermal power and the particle acceleration strength driven by the low-coronal reconnection event, the novel law derives a connection between the low coronal and IP particle acceleration driven by the reconnection and the CME events.
{ Previous attempts to identify GBR-like scaling laws connecting \lx\ and \lr\ in the metric band using stellar data had not been successful. This study provides a philosophy and approach to extend an \lx-\lr\ relationship for long wavelengths by taking the emission drivers into account.}
Additionally, we derived scaling laws linking \lx\ and \rec\ to \Pc. 
{ By combining the derived and known scaling laws with disk-integrated \lx\ and \lr\ observable in stellar events, \vcme\  and  \rec\ can be estimated. Additionally, assuming a plasma emission mechanism for type II bursts, the frequency drift rate can be used to estimate $\mathrm{\nabla_h n_e}$ in the corona, once \vcme\ is estimated. 
The presented study takes a top-down approach to understanding the interconnections between the two major sub-events of the flare-CME-type II phenomena. There needs to be a bottom-up approach using modeling to understand the emergence of the various scaling laws, particularly the \Pf-\Pc\ scaling.} 

\begin{acknowledgements}
AM, NG and HR are partly supported by NASA's STEREO project and LWS program. SA was partially supported by NSF grant, AGS-2043131. We thank the CDAW team for maintaining an up-to-date catalog\footnote{\href{https://cdaw.gsfc.nasa.gov/CME_list/}{https://cdaw.gsfc.nasa.gov/CME\_list/}} of solar CMEs detected by the Large Angle and Spectrometric Coronagraph (LASCO) on board the Solar and Heliospheric Observatory (SOHO) mission.  
AM acknowledges Pertti M\"{a}kel\"{a} for maintaining an up-to-date DH type II catalog. AM acknowledges Vratislav Krupar and NASA's Space Physics Data Facility (SPDF)\footnote{\href{https://spdf.gsfc.nasa.gov/}{https://spdf.gsfc.nasa.gov/}} for the calibrated radio dynamic spectra.
AM acknowledges the developers of the various Python modules namely Numpy \citep{numpy}, Astropy \citep{astropy}, Matplotlib \citep{matplotlib}, and Pandas~\citep{reback20_pandas}.
\end{acknowledgements}

%





\bibliographystyle{aa}
\bibliography{paper}

\begin{thebibliography}{38}
\expandafter\ifx\csname natexlab\endcsname\relax\def\natexlab#1{#1}\fi

\bibitem[{{Airapetian} \& {Holman}(1998)}]{airapetian98_GBRtheory}
{Airapetian}, V.~S. \& {Holman}, G.~D. 1998, \apj, 501, 805

\bibitem[{{Alvarado-G{\'o}mez} {et~al.}(2022){Alvarado-G{\'o}mez}, {Drake}, {Cohen}, {Fraschetti}, {Garraffo}, \& {Poppenh{\"a}ger}}]{alvarado22_CME_star}
{Alvarado-G{\'o}mez}, J.~D., {Drake}, J.~J., {Cohen}, O., {et~al.} 2022, Astronomische Nachrichten, 343, e10100

\bibitem[{{Astropy Collaboration} {et~al.}(2013){Astropy Collaboration}, {Robitaille}, {Tollerud}, {Greenfield}, {Droettboom}, {Bray}, {Aldcroft}, {Davis}, {Ginsburg}, {Price-Whelan}, {Kerzendorf}, {Conley}, {Crighton}, {Barbary}, {Muna}, {Ferguson}, {Grollier}, {Parikh}, {Nair}, {Unther}, {Deil}, {Woillez}, {Conseil}, {Kramer}, {Turner}, {Singer}, {Fox}, {Weaver}, {Zabalza}, {Edwards}, {Azalee Bostroem}, {Burke}, {Casey}, {Crawford}, {Dencheva}, {Ely}, {Jenness}, {Labrie}, {Lim}, {Pierfederici}, {Pontzen}, {Ptak}, {Refsdal}, {Servillat}, \& {Streicher}}]{astropy}
{Astropy Collaboration}, {Robitaille}, T.~P., {Tollerud}, E.~J., {et~al.} 2013, \aap, 558, A33

\bibitem[{{Callingham} {et~al.}(2021){Callingham}, {Vedantham}, {Shimwell}, {Pope}, {Davis}, {Best}, {Hardcastle}, {R{\"o}ttgering}, {Sabater}, {Tasse}, {van Weeren}, {Williams}, {Zarka}, {de Gasperin}, \& {Drabent}}]{callingham21_noGBR_LoTSS}
{Callingham}, J.~R., {Vedantham}, H.~K., {Shimwell}, T.~W., {et~al.} 2021, Nature Astronomy, 5, 1233

\bibitem[{{Crosley} \& {Osten}(2018)}]{2018ApJ...856...39C}
{Crosley}, M.~K. \& {Osten}, R.~A. 2018, \apj, 856, 39

\bibitem[{{Ginzburg} \& {Zhelezniakov}(1958)}]{ginzburg1958}
{Ginzburg}, V.~L. \& {Zhelezniakov}, V.~V. 1958, \sovast, 2, 653

\bibitem[{Gopalswamy(2006)}]{gopal06_CMEtypeII_book}
Gopalswamy, N. 2006, Coronal Mass Ejections and Type II Radio Bursts (American Geophysical Union (AGU)), 207--220

\bibitem[{{Gopalswamy}(2011)}]{Gopal11_PREconf}
{Gopalswamy}, N. 2011, in Planetary, Solar and Heliospheric Radio Emissions (PRE VII), ed. H.~O. {Rucker}, W.~S. {Kurth}, P.~{Louarn}, \& G.~{Fischer}, 325--342

\bibitem[{{Gopalswamy} {et~al.}(2005){Gopalswamy}, {Aguilar-Rodriguez}, {Yashiro}, {Nunes}, {Kaiser}, \& {Howard}}]{gopal05_typeIIs_SEPs}
{Gopalswamy}, N., {Aguilar-Rodriguez}, E., {Yashiro}, S., {et~al.} 2005, Journal of Geophysical Research (Space Physics), 110, A12S07

\bibitem[{Gopalswamy {et~al.}(2017)Gopalswamy, Akiyama, Yashiro, \& Xie}]{Gopal17_FREDtech}
Gopalswamy, N., Akiyama, S., Yashiro, S., \& Xie, H. 2017, Proceedings of the International Astronomical Union, 13, 258–262

\bibitem[{{Gopalswamy} {et~al.}(2018{\natexlab{a}}){Gopalswamy}, {Akiyama}, {Yashiro}, \& {Xie}}]{gopal18_coronalRcflx_IPburstconnection}
{Gopalswamy}, N., {Akiyama}, S., {Yashiro}, S., \& {Xie}, H. 2018{\natexlab{a}}, Journal of Atmospheric and Solar-Terrestrial Physics, 180, 35

\bibitem[{{Gopalswamy} {et~al.}(2019){Gopalswamy}, {M{\"a}kel{\"a}}, \& {Yashiro}}]{Gopal19_DHtypeII_catalog}
{Gopalswamy}, N., {M{\"a}kel{\"a}}, P., \& {Yashiro}, S. 2019, Sun and Geosphere, 14, 111

\bibitem[{{Gopalswamy} {et~al.}(2018{\natexlab{b}}){Gopalswamy}, {M{\"a}kel{\"a}}, {Yashiro}, {Lara}, {Xie}, {Akiyama}, \& {MacDowall}}]{gopal18_SGRE-typeIIlink}
{Gopalswamy}, N., {M{\"a}kel{\"a}}, P., {Yashiro}, S., {et~al.} 2018{\natexlab{b}}, \apjl, 868, L19

\bibitem[{{Gudel} {et~al.}(1993){Gudel}, {Schmitt}, {Bookbinder}, \& {Fleming}}]{guedel93_dM_GBZ}
{Gudel}, M., {Schmitt}, J. H.~M.~M., {Bookbinder}, J.~A., \& {Fleming}, T.~A. 1993, \apj, 415, 236

\bibitem[{{Guedel} \& {Benz}(1993)}]{guedel93_GBR_F-M}
{Guedel}, M. \& {Benz}, A.~O. 1993, \apjl, 405, L63

\bibitem[{{Guedel} {et~al.}(1995){Guedel}, {Schmitt}, \& {Benz}}]{guedel95_GBR_FGgiants}
{Guedel}, M., {Schmitt}, J.~H.~M.~M., \& {Benz}, A.~O. 1995, \aap, 302, 775

\bibitem[{Harris {et~al.}(2020)Harris, Millman, van~der Walt, Gommers, Virtanen, Cournapeau, Wieser, Taylor, Berg, Smith, Kern, Picus, Hoyer, van Kerkwijk, Brett, Haldane, del R{'{\i}}o, Wiebe, Peterson, G{'{e}}rard-Marchant, Sheppard, Reddy, Weckesser, Abbasi, Gohlke, \& Oliphant}]{numpy}
Harris, C.~R., Millman, K.~J., van~der Walt, S.~J., {et~al.} 2020, Nature, 585, 357

\bibitem[{Hastie {et~al.}(2009)Hastie, Tibshirani, \& Friedman}]{ML_book}
Hastie, T., Tibshirani, R., \& Friedman, J. 2009, The Elements of Statistical Learning: Data Mining, Inference, and Prediction, Springer series in statistics (Springer)

\bibitem[{{Howard} {et~al.}(2023){Howard}, {Vourlidas}, \& {Stenborg}}]{howard23_CME_review}
{Howard}, R.~A., {Vourlidas}, A., \& {Stenborg}, G. 2023, Frontiers in Astronomy and Space Sciences, 10, 1264226

\bibitem[{Hunter(2007)}]{matplotlib}
Hunter, J.~D. 2007, Computing in science \& engineering, 9, 90

\bibitem[{{Kazachenko}(2023)}]{kazachenko23_Ribbondb}
{Kazachenko}, M.~D. 2023, \apj, 958, 104

\bibitem[{{Kazachenko} {et~al.}(2017){Kazachenko}, {Lynch}, {Welsch}, \& {Sun}}]{Kazachenko17_Ribbondb}
{Kazachenko}, M.~D., {Lynch}, B.~J., {Welsch}, B.~T., \& {Sun}, X. 2017, \apj, 845, 49

\bibitem[{{Kumari} {et~al.}(2023){Kumari}, {Morosan}, {Kilpua}, \& {Daei}}]{anshu23_typeII_stats}
{Kumari}, A., {Morosan}, D.~E., {Kilpua}, E.~K.~J., \& {Daei}, F. 2023, \aap, 675, A102

\bibitem[{McLean \& Labrum(1985)}]{Mclean85_book}
McLean, D. \& Labrum, N. 1985, {Solar Radio Astrophysics} (Cambridge University Press)

\bibitem[{{Melrose}(1970)}]{melrose1970}
{Melrose}, D.~B. 1970, Australian Journal of Physics, 23, 885

\bibitem[{{Miteva} {et~al.}(2017){Miteva}, {Samwel}, \& {Krupar}}]{Miteva17_SEP-radburst_link}
{Miteva}, R., {Samwel}, S.~W., \& {Krupar}, V. 2017, Journal of Space Weather and Space Climate, 7, A37

\bibitem[{{Mohan} {et~al.}(2024{\natexlab{a}}){Mohan}, {Gopalswamy}, {Kumari}, {Akiyama}, \& {G}}]{Atul24_DHtypeIVcatalog}
{Mohan}, A., {Gopalswamy}, N., {Kumari}, A., {Akiyama}, S., \& {G}, S. 2024{\natexlab{a}}, \apj, 971, 86

\bibitem[{{Mohan} {et~al.}(2024{\natexlab{b}}){Mohan}, {Mondal}, {Wedemeyer}, \& {Gopalswamy}}]{atul24_ADLeotypeIV}
{Mohan}, A., {Mondal}, S., {Wedemeyer}, S., \& {Gopalswamy}, N. 2024{\natexlab{b}}, \aap, 686, A51

\bibitem[{{Osten} \& {Bastian}(2008)}]{Osten08_ADLeoFinebursts}
{Osten}, R.~A. \& {Bastian}, T.~S. 2008, \apj, 674, 1078

\bibitem[{pandas~development team(2020)}]{reback20_pandas}
pandas~development team, T. 2020, pandas-dev/pandas: Pandas

\bibitem[{{Pevtsov} {et~al.}(2003){Pevtsov}, {Fisher}, {Acton}, {Longcope}, {Johns-Krull}, {Kankelborg}, \& {Metcalf}}]{petsov03_Lx_recflx_solar-stellar}
{Pevtsov}, A.~A., {Fisher}, G.~H., {Acton}, L.~W., {et~al.} 2003, \apj, 598, 1387

\bibitem[{Sindhuja \& Gopalswamy(2020)}]{Sindhuja20_Recflx_xflr}
Sindhuja, G. \& Gopalswamy, N. 2020, The Astrophysical Journal, 889, 104

\bibitem[{{Villadsen} \& {Hallinan}(2019)}]{2019ApJ...871..214V}
{Villadsen}, J. \& {Hallinan}, G. 2019, \apj, 871, 214

\bibitem[{{Villadsen} {et~al.}(2014){Villadsen}, {Hallinan}, {Bourke}, {G{\"u}del}, \& {Rupen}}]{Villadsen14_First_detect_SLS_inRadio_VLA}
{Villadsen}, J., {Hallinan}, G., {Bourke}, S., {G{\"u}del}, M., \& {Rupen}, M. 2014, \apj, 788, 112

\bibitem[{{Wild}(1970)}]{wild1970}
{Wild}, J.~P. 1970, Proceedings of the Astronomical Society of Australia, 1, 365

\bibitem[{{Wild} \& {McCready}(1950)}]{wild50_bursttypes}
{Wild}, J.~P. \& {McCready}, L.~L. 1950, Australian Journal of Scientific Research A Physical Sciences, 3, 387

\bibitem[{{Yiu} {et~al.}(2024){Yiu}, {Vedantham}, {Callingham}, \& {G{\"u}nther}}]{Yiu24_noGBR_vlass}
{Yiu}, T.~W.~H., {Vedantham}, H.~K., {Callingham}, J.~R., \& {G{\"u}nther}, M.~N. 2024, \aap, 684, A3

\bibitem[{{Zic} {et~al.}(2020){Zic}, {Murphy}, {Lynch}, {Heald}, {Lenc}, {Kaplan}, {Cairns}, {Coward}, {Gendre}, {Johnston}, {MacGregor}, {Price}, \& {Wheatland}}]{Zic20_typeIV_ProximaCen}
{Zic}, A., {Murphy}, T., {Lynch}, C., {et~al.} 2020, \apj, 905, 23

\end{thebibliography}



\end{document}